\documentclass[aps,prl,floatfix,twocolumn,showpacs,superscriptaddress]{revtex4-2}  
\usepackage{graphicx}  
\usepackage{amsmath,amssymb}

\usepackage{physics}

\usepackage{flushend}
\usepackage[colorlinks = true,
            linkcolor = blue,
            urlcolor  = blue,
            citecolor = blue,
            anchorcolor = blue]{hyperref}
\usepackage{array}
\usepackage{float}
\usepackage{natbib}
\usepackage{booktabs}
\usepackage{color}
\usepackage{times}

\begin{document}

\title{Quantum metric senses a persistent spin helix}

\author{Awadhesh Narayan} 
\email[]{awadhesh@iisc.ac.in} 
\affiliation{Solid State and Structural Chemistry Unit, Indian Institute of Science, Bangalore 560012, India}
\affiliation{Materials Theory, ETH Zurich, Wolfgang-Pauli-Strasse 27, CH 8093 Zurich, Switzerland}

\begin{abstract}
Persistent spin helices are a manifestation of symmetry-protected spin textures in systems with balanced spin-orbit coupling. They enable long-lived spin structures that are of interest for spintronics and coherent spin manipulation. The quantum metric has recently emerged as a promising tool for characterizing the geometric structure of quantum states. Here, we demonstrate that the quantum metric provides a sensitive geometric probe of the persistent spin helix. Within the Rashba-Dresselhaus Hamiltonian, we analytically evaluate the quantum metric components and uncover a divergent geometric contribution that emerges precisely at the persistent spin helix condition. We reveal that this divergence originates from a hidden line degeneracy that forms when the strengths of Rashba and Dresselhaus spin–orbit coupling become equal. We further study the role of higher-order cubic spin-orbit interactions and determine how these corrections regularize the geometric response and control the scaling behavior of the quantum metric. Our results establish quantum geometry as a powerful framework for identifying and characterizing persistent spin helices and related symmetry-protected spin textures.
\end{abstract}

\maketitle

\date{\today}

\begin{figure}[b]
    \centering
    \includegraphics[width=0.40\textwidth]{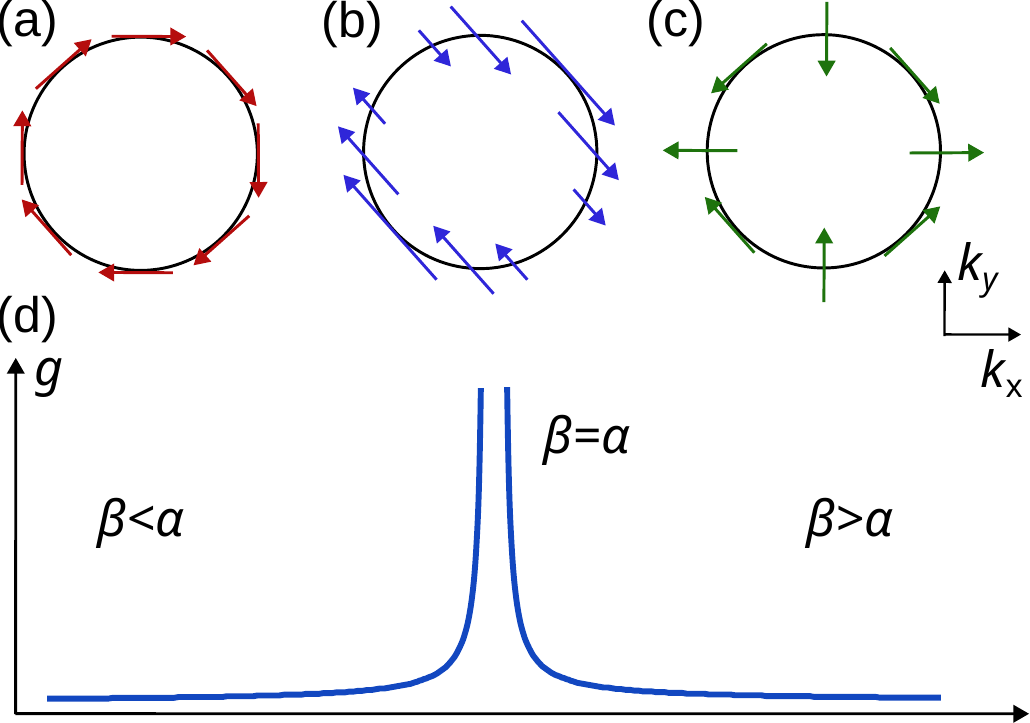}
    \caption{\textbf{Spin texture and variation of the quantum metric.} Schematic of the spin texture with (a) Rashba ($\alpha \neq 0, \beta=0$), (b) persistent spin helix ($\alpha=\beta$), and (c) Dresselhaus ($\alpha=0, \beta \neq 0$) terms plotted in the $k_x-k_y$ plane. (d) Schematic of the quantum metric components, $g_{\mu\nu}$, with varying Dresselhaus to Rashba strength ratio, $\beta/\alpha$. We discover a sharp enhancement of all components of the metric when the Rashba and Dresselhaus strengths become equal, $\beta=\alpha$. This is a striking signature of the persistent spin helix.}\label{schematic}
\end{figure}

\textit{Introduction--} Spin-orbit coupling plays an important role in diverse quantum phenomena. The interplay of Rashba~\cite{bychkov1984oscillatory} and Dresselhaus~\cite{dresselhaus1955spin} types of spin-orbit coupling has been a fruitful arena for exploring band topology, magnetism, and ferroelectricity, among other features~\cite{galitski2013spin,picozzi2014ferroelectric,manchon2015new}. When the strengths of the Rashba and Dresselhaus spin-orbit coupling become equal, a special spin texture called the persistent spin helix emerges [Fig.~\ref{schematic}(a)-(c)]~\cite{bernevig2006exact,schliemann2003nonballistic,koralek2009emergence}. Under this condition, the spin–orbit Hamiltonian acquires an emergent SU(2) symmetry, which protects the helical spin-density wave from rapid spin relaxation. As a result, spins exhibit an unusually long lifetime compared with generic spin textures~\cite{koralek2009emergence,walser2012direct,sasaki2014direct}. Since its original discovery, the persistent spin helix has become an important platform for studying symmetry-protected spin transport~\cite{schliemann2017colloquium,kohda2017physics}. Beyond initial explorations in semiconductor quantum wells, this concept has been extended to a number of material platforms~\cite{tao2018persistent,autieri2019persistent,tao2021perspectives,narayan2015class,ji2022symmetry,lu2023strain}.

The quantum metric is a fundamental geometric quantity that characterizes the distance between nearby quantum states in parameter space, such as crystal momentum in Bloch bands~\cite{provost1980riemannian}. It forms the real part of the quantum geometric tensor, whose imaginary part is the celebrated Berry curvature~\cite{cheng2010quantum,rossi2021quantum,torma2023essay}. Recently, the quantum metric has come to fore in condensed matter systems, where it plays a central role in phenomena such as superfluid weight in flat bands, optical responses, and transport properties of topological materials~\cite{gao2023quantum,wang2023quantum,sala2025quantum,liu2025quantum,yu2025quantum}. Going beyond indirect effects, very recently, there have been notable experiments on the direct measurement of the quantum metric~\cite{yu2020experimental,kang2025measurements,kim2025direct}. We note that early works have suggested the quantum metric (also termed fidelity tensor) as a probe of phase transitions~\cite{ma2010abelian,wang2015fidelity}.

In this contribution, we propose and demonstrate the quantum metric as a sensitive probe of the persistent spin helix. Within the Rashba–Dresselhaus Hamiltonian, we analytically evaluate the quantum metric components and uncover a divergent geometric contribution that emerges precisely at the persistent spin helix condition [Fig.~\ref{schematic}(d)]. We show that this divergence originates from a hidden line degeneracy appearing when the strengths of the Rashba and Dresselhaus spin–orbit coupling become equal, revealing a direct link between band degeneracies and singular features of the quantum geometry. We further investigate the effect of higher-order cubic spin–orbit terms and determine how these corrections regulate the geometric response and control the scaling of the quantum metric. Our results highlight quantum geometry as a promising framework for identifying and characterizing persistent spin helix states and, more broadly, emphasize the utility of quantum geometric diagnostics for uncovering symmetry-protected structures in spin–orbit coupled materials.

\begin{figure}[t]
    \centering
    \includegraphics[width=0.5\textwidth]{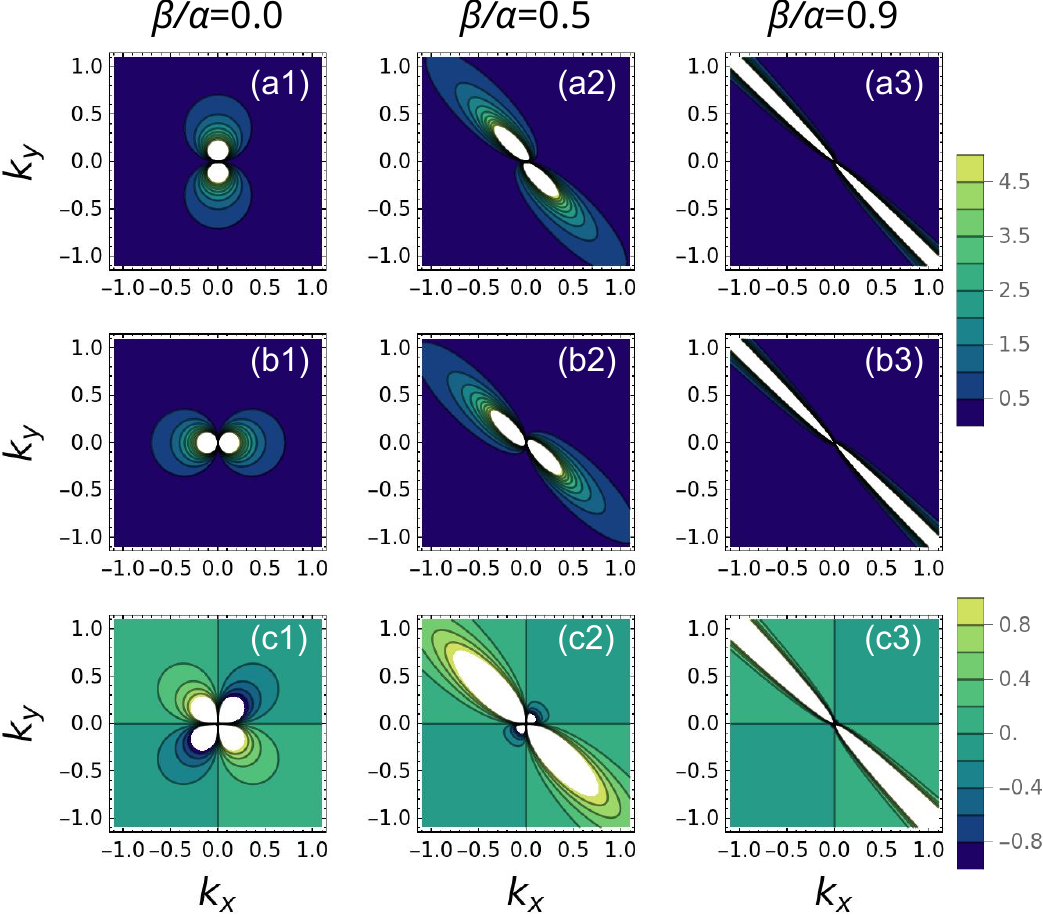}
    \caption{\textbf{Distribution of quantum metric components.} The momentum space distribution of the quantum metric components (a1)-(a3) $g_{xx}(\mathbf{k})$, (b1)-(b3) $g_{yy}(\mathbf{k})$, and (c1)-(c3) $g_{xy}(\mathbf{k})$ for different Dresselhaus to Rashba strength ratios, $\beta/\alpha$, noted at the top of each column. In the absence of the Dresselhaus term, the diagonal components $g_{xx}(\mathbf{k})$ and $g_{yy}(\mathbf{k})$ present mirror symmetries along $k_y=0$ and $k_x=0$, respectively, as shown in panels (a1) and (b1). The off-diagonal component, $g_{xy}(\mathbf{k})$, exhibits a quadrupolar petal-like pattern with alternating equal-sized positive and negative lobes, as shown in panel (c1). As the $\beta/\alpha$ ratio increases, the mirror symmetry in the diagonal elements is lost [(a2)-(b2)]. The positive and negative lobes of the off-diagonal component become unequal [(c2)]. The distributions become skewed along the $k_x=-k_y$ direction. As the $\beta/\alpha$ ratio approaches one, a singular ridge forms along the $k_x=-k_y$ direction, where all components of the quantum metric take large values, as shown in panels (a3)-(c3). This enhancement, together with the line degeneracy having its direction locked by symmetry to the spin texture, signals the formation of the persistent spin helix.}\label{g_distribution}
\end{figure}

\textit{Model--} We consider the Rashba-Dresselhaus Hamiltonian given by~\cite{bernevig2006exact}

\begin{equation}
    H=\frac{\hbar^2k^2}{2m}I+\alpha(k_y\sigma_x-k_x\sigma_y)+\beta(k_x\sigma_x-k_y\sigma_y).
\end{equation}

Here $m$ is the effective mass, $(k_x,k_y)$ are the in-plane momenta, and $\alpha$ and $\beta$ are the Rashba and Dresselhaus spin-orbit coupling strengths, respectively. The Pauli matrices are denoted by $\sigma_i$ with $i=x,y,z$, and $I$ is the $2\times2$ identity matrix. We set $\hbar=1$. Writing the Hamiltonian as $H=d_{0}(\mathbf{k})I+\mathbf{d}(\mathbf{k})\cdot\mathbf{\sigma}$, the energy eigenvalues are $E_{\pm}=d_{0}(\mathbf{k})\pm|\mathbf{d}(\mathbf{k})|$.

\textit{Results and discussion--} We begin by calculating the quantum metric components, $g_{\mu\nu}(\mathbf{k})$, which are given by $g_{\mu\nu}^{\pm}(\mathbf{k})=\frac{1}{4}\partial_{k_{\mu}}\hat{\mathbf{d}}\cdot\partial_{k_{\nu}}\hat{\mathbf{d}}$~\cite{cheng2010quantum}. Here $\hat{\mathbf{d}}=\mathbf{d}(\mathbf{k})/|\mathbf{d}(\mathbf{k})|$, $\mu,\nu\in (x,y)$ and $\pm$ denotes the band index. The quantum metric components for the lower band read

\begin{align}
    g_{xx}(\mathbf{k})=\frac{k_y^2(\alpha^2-\beta^2)^2}{4[4k_xk_y\alpha\beta+(\alpha^2+\beta^2)(k_x^2+k_y^2)]^2}, \nonumber \\
    g_{yy}(\mathbf{k})=\frac{k_x^2(\alpha^2-\beta^2)^2}{4[4k_xk_y\alpha\beta+(\alpha^2+\beta^2)(k_x^2+k_y^2)]^2}, \nonumber \\
    g_{xy}(\mathbf{k})= -\frac{k_xk_y(\alpha^2-\beta^2)^2}{4[4k_xk_y\alpha\beta+(\alpha^2+\beta^2)(k_x^2+k_y^2)]^2}.
\end{align}

At a first glance, it appears that all the components vanish at the persistent spin helix condition $\alpha=\beta$. However, a more careful evaluation is required due to a subtle singularity appearing in the denominator. We note that the Berry curvature vanishes for all choices of $\alpha$ and $\beta$ for our Rashba-Dresselhaus model.

To understand the behavior of $g(\mathbf{k})$, we introduce the difference between the Rashba and Dresselhaus spin-orbit coupling strengths $\delta=\alpha-\beta$ and rotated coordinates $k_{\pm}=(k_x\pm k_y)/\sqrt{2}$. We find the scaling of the quantum metric component in the vicinity of the persistent spin helix condition as, $g_{xx}(\mathbf{k})\sim \frac{\delta^2}{\beta^2k_+^4}$. This diverges strongly as $k_+\rightarrow 0$, i.e., along the $k_x=-k_y$ direction in the momentum plane. We note that this is precisely the direction of the persistent spin texture [Fig.~\ref{schematic}(b)], hinting towards an underlying connection between the quantum metric and the persistent spin helix.

\begin{figure*}[t]
    \centering
    \includegraphics[width=0.7\textwidth]{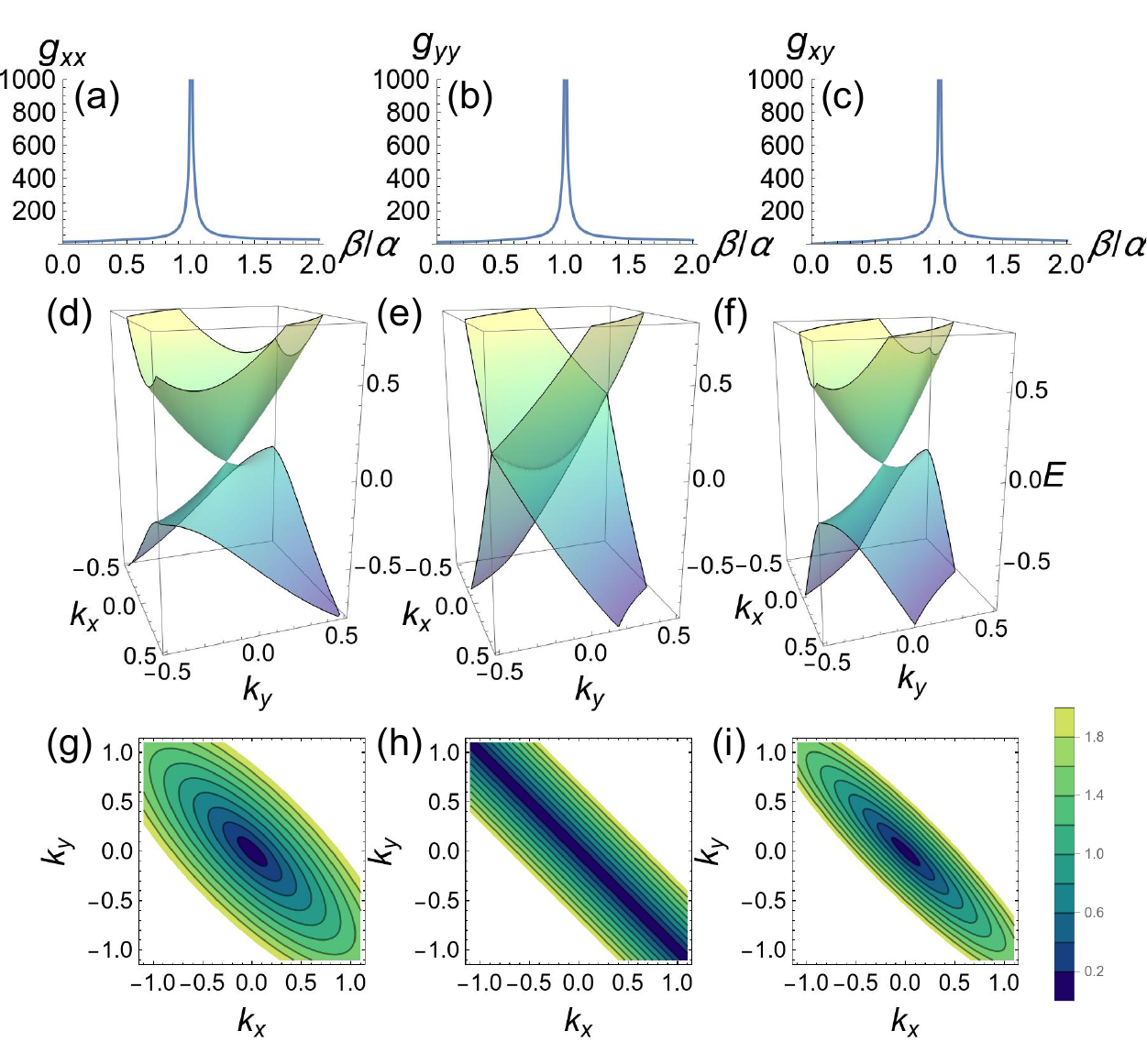}
    \caption{\textbf{Integrated quantum metric and nature of band structures.} Quantum metric components (a) $g_{xx}$, (b) $g_{yy}$, and (c) $g_{xy}$ integrated over the momentum space, plotted as a function of the Dresselhaus to Rashba strength ratio, $\beta/\alpha$. We find a sharp enhancement in the quantum metric components near the persistent spin helix condition ($\alpha=\beta$). The band structure for the Rashba-Dresselhaus model for (d) $\beta/\alpha=0.5$, (e) $\beta/\alpha=1.0$, and (f) $\beta/\alpha=1.5$. The corresponding band gaps are presented in panels (g)-(i). Notably a hidden line degeneracy, with a vanishing band gap, appears along the $k_x=-k_y$ direction when the persistent spin helix forms, as shown in panels (e) and (h).}\label{g_integrated_bands}
\end{figure*}

To gain further insights, we next examine the momentum space distribution of the quantum metric components as the ratio of the Dresselhaus to Rashba spin-orbit coupling strengths, $\beta/\alpha$, is varied. Fig.~\ref{g_distribution} shows the momentum-space distribution of the quantum metric tensor components, $g_{\mu\nu}(\mathbf{k})$ for several values of the Dresselhaus to Rashba strength ratio, $\beta/\alpha$. In the pure Rashba limit ($\beta/\alpha=0$), the systems exhibits a symmetric structure of the quantum metric [Fig.~\ref{g_distribution}(a1)-(c1)]. The diagonal components, $g_{xx}(\mathbf{k})$ and $g_{yy}(\mathbf{k})$, display mirror symmetries with respect to the $k_y=0$ and $k_x=0$ axes, respectively. The off-diagonal component, $g_{xy}(\mathbf{k})$, exhibits a quadrupolar distribution with alternating positive and negative lobes, which are of equal size. As the Dresselhaus coupling is introduced, the metric distributions become anisotropic and skewed in momentum space. In particular, the diagonal components lose their simple mirror symmetry, and the off-diagonal component develops unequal lobes, indicating an imbalance in the geometric response along different momentum directions [Fig.~\ref{g_distribution}(a2)-(c2)]. A striking feature appears as the system approaches the persistent spin helix condition $\beta/\alpha\rightarrow 1$, shown in Fig.~\ref{g_distribution}(a3)-(c3). A sharp ridge-like feature emerges along $k_+\rightarrow 0$. This ridge corresponds to a strong enhancement of the local quantum metric near the persistent spin helix condition. All components of the quantum metric take large values along this $k_x=-k_y$ direction. Notably, this agrees well with the analytic scaling we found previously. For $\beta/\alpha>1$, the ridge disappears, giving way to strongly asymmetric quantum metric distributions as the system moves away from the persistent spin helix regime.

The integrated quantum metric components, obtained by integrating the momentum-resolved quantities $g_{\mu\nu}(\mathbf{k})$ over the momentum space, are shown in Fig.~\ref{g_integrated_bands}(a)-(c), as a function of the Dresselhaus to Rashba spin–orbit coupling ratio $\beta/\alpha$. As we found, the momentum space distributions of the quantum metric components develop a pronounced ridge along the $k_x=-k_y$ direction as the ratio $\beta/\alpha$ approaches unity. Upon integration over momentum space, this localized enhancement manifests as a sharp peak in the integrated all the quantum metric components near $\alpha=\beta$. We can gain insight into the nature of the divergence at the persistent spin helix by examining the integrated quantum metric. By introducing appropriate low and high momentum cutoffs, we find $g_{xx}\sim \frac{\beta}{|\delta|}\ln\frac{\Lambda_{UV}}{\Lambda_{IR}}$. Here $\delta=\alpha-\beta$ is the difference between the Rashba and Dresselhaus strengths, and $\Lambda_{IR}$ and $\Lambda_{UV}$ are the low and high momentum cutoffs. We find that the quantum metric components diverge inversely with $\delta=\alpha-\beta$, and this divergence is symmetric about $\alpha=\beta$. Indeed, this is also borne out from our numerical calculations shown in Fig.~\ref{g_integrated_bands}(a)-(c).

To elucidate the origin of this pronounced enhancement in the quantum metric for the persistent spin helix, we next analyze the band structures of the Rashba-Dresselhaus model for representative Dresselhaus to Rashba spin–orbit coupling ratios. The band structures are shown in Fig.~\ref{g_integrated_bands}(d)-(f) for $\beta/\alpha=$0.5, 1, and 1.5, while the Fig.~\ref{g_integrated_bands}(g)-(i) display the corresponding momentum-dependent band gaps. For $\beta/\alpha\neq 1$, the bands remain separated by a finite gap throughout momentum space except at a single point, namely the origin $\mathbf{k}=0$. This results in moderate but tunable values of the quantum metric~\cite{luo2025tunable,reja2025field}. In striking contrast, when the persistent spin helix condition $\beta/\alpha=1$ is met, the band gap collapses along the line $k_x=-k_y$, giving rise to a hidden line degeneracy. We recall that this is precisely the direction of the persistent spin texture as well as the region with the sharp ridge in the momentum space distribution of the quantum metric components. This line degeneracy produces a strong enhancement of the quantum metric near the nodal direction. Consequently, the integrated metric exhibits a pronounced peak at $\alpha=\beta$, providing a direct geometric signature of the underlying persistent spin helix. 

We further note that at $\alpha=\beta$, the spin-orbit field becomes unidirectional resulting in an emergent SU(2) symmetry~\cite{bernevig2006exact}. This line degeneracy is the momentum-space manifestation of the spectral mapping identified by Bernevig and coauthors~\cite{bernevig2006exact}. As the degeneracy is protected by this emergent SU(2) symmetry, which Ref.~\cite{bernevig2006exact} showed is stable against density-density interactions, we expect the nodal line and hence the qualitative divergence structure of the metric to be robust. Interactions can renormalize the effective spin-orbit strengths and mass, shifting the location and magnitude of the enhancement but not removing it, as long as the symmetry is preserved. Furthermore, in three-dimensional non-centrosymmetric bulk with coexisting Rashba and Dresselhaus terms, the balanced condition promotes the nodal line to a nodal structure in the three-dimensional Brillouin zone. The quantum metric enhancement inherits the corresponding codimension.

Till now we have carried out our analysis using linear-in-momentum spin-orbit coupling terms in the Rashba-Dresselhaus Hamiltonian. In general, higher-order corrections to the spin-orbit coupling can be present. One of the most important is the cubic Dresselhaus correction. We next assess the robustness of the quantum geometric enhancement that we found near the persistent spin helix condition, to such a term. The third-order Dresselhaus correction term to the Hamiltonian reads~\cite{koralek2009emergence}

\begin{equation}
    H'=-\beta_3(k_xk_y^2\sigma_x-k_yk_x^2\sigma_y).
\end{equation}

Without the cubic correction, at the persistent spin helix condition $\alpha=\beta$, the linear spin–orbit fields "cancel" along $k_x=-k_y$ leading to a line degeneracy. However, the cubic contribution produces a finite band splitting that scales as $\beta_3k^3$, lifting the line degeneracy. This leads to a qualitatively different nature of the quantum metric in comparison to the linear Rashba-Dresselhaus spin-orbit coupling terms. We present the complete expressions for the energy eigenvalues in the presence of the cubic Dresselhaus term in the supplement~\cite{supplement}.

Since the quantum metric typically scales inversely with the square of the band gap, the cubic Dresselhaus term introduces a natural infrared cutoff to the quantum metric singularity (see supplement for full expressions of the quantum metric components in presence of this term~\cite{supplement}). We find that the cubic Dresselhaus term controls the magnitude of the geometric response near the persistent spin helix point. Thus, while the linear Rashba–Dresselhaus model predicts a divergent quantum metric, the inclusion of realistic higher-order spin-orbit coupling terms converts this divergence into a large but finite enhancement of the quantum metric. While the precise values of $\beta_3$ depend upon the carrier densities and Fermi wave vectors, typically, the cubic term is an order of magnitude smaller than the linear Dresselhaus term. Therefore, the enhancement in the quantum metric at the persistent spin helix condition, is quite substantial and should be directly accessible to experiments.

Nonlinear transport experiments involve the band-normalized quantum metric or the Berry connection polarizability tensor~\cite{gao2014field,gao2019semiclassical,kaplan2023unifying,kaplan2024unification,gao2025quantum}. We find that for our two-band Rashba-Dresselhaus model the band-normalized metric tensor reduces exactly to $G_{\mu\nu} = -g_{\mu\nu}/|\mathbf{d}|$, i.e. the quantum metric divided by one power of the gap. See supplement for its closed form expressions~\cite{supplement}. From the Fermi-surface projection of this quantity, which is directly relevant for nonlinear transport experiments, we find that at the persistent spin helix condition, the Fermi contour crosses the nodal direction $k_x=-k_y$. The band-normalized metric is therefore \emph{more} strongly enhanced than the integrated bare metric. The cubic Dresselhaus term regularizes this enhancement, exactly as it regularizes the integrated quantum metric. Along the former nodal line the gap acquires a value $|\mathbf{d}| = \tfrac{1}{2}\beta_3 |k_-|^3$, and the Fermi surface projected band-normalized metric develops a finite peak at $\alpha = \beta$. This peak value scales as $G_{\mathrm{tr}} \sim \frac{\beta}{\beta_3^{2}}$, where $\mathrm{tr}$ denotes the trace of the Berry connection polarizability tensor.

\textit{Outlook and summary--} Our predictions can be tested using state-of-the-art nonlinear transport experiments~\cite{sala2025quantum}, where gating or doping asymmetry can tune the ratio of Dresselhaus to Rashba spin–orbit coupling~\cite{koralek2009emergence}. The integrated quantum metric is proportional to the quantum weight~\cite{PhysRevResearch.7.023158}, which is fixed by an optical sum rule and is accessible through probes such as x-ray scattering or electron energy-loss spectroscopy, rather than through nonlinear transport. These experimental platforms can enable the detection of the predicted sharp enhancement near the persistent spin helix condition. Moreover, the effective two-level structure underlying our analysis simplifies the quantum metric response, making the predicted geometric enhancement particularly amenable to direct experimental measurement~\cite{kim2025direct}. 

In summary, we showed that the quantum metric provides a sensitive probe of the persistent spin helix in Rashba–Dresselhaus systems. An analytic evaluation revealed a divergent geometric contribution at the persistent spin helix condition, originating from a hidden line degeneracy that appears when the Rashba and Dresselhaus strengths become equal. We further demonstrated that cubic spin–orbit corrections lift this degeneracy, thereby regularizing the divergence and controlling the scaling of the quantum metric. Our results establish quantum geometry as a powerful framework for diagnosing persistent spin helices and related symmetry-protected spin structures in spin–orbit coupled materials.

\textit{Note added--} A related preprint appeared recently~\cite{chakraborti2026probing}.

\textit{Acknowledgments--}
I thank Nicola Spaldin for insightful discussions. I acknowledge the Swiss National Science Foundation for sabbatical support and Materials Theory, ETH Zurich for the kind hospitality. I also thank Department of Science and Technology, Core Research Grant (CRG/2023/000114) for funding.

\bibliography{references}

\end{document}